\documentstyle[11pt]{article}
\newfont{\ssq}{cmss10}
\begin{document}
\begin{center}
{\bf Comment on ``Two-dimensional charged-exciton complexes''
by Varga, K -- Reply}
\bigskip
\bigskip
\end{center}   

\begin{center}
A. Thilagam  \\
Faculty of Science \\ 
Northern Territory University \\
Darwin NT 0909 \\
Australia
\end{center}

\bigskip

\begin{abstract}
We respond to   criticisms raised by K. Varga  (cond-mat/9802262)
and reaffirm that the results in our
original paper obtained using a two-body analytical method
remains valid within the framework of an effective 
excitonic composite model. The conceptual model of the excitonic 
systems as well as the numerical method based on variational
functions utilized by Varga differ significantly from ours.
Hence comparison of binding energies of the
charged-biexciton remains questionable. 
In this reply, we discuss the shortcomings of modelling
the charged-biexciton as a  five-body system and  treating
excitonic complexes as atomic systems, as
done in Varga's Comment. We also clarify the somewhat 
misleading statement that the "charged-biexciton complex is 
stable for any value of the mass ratio" in our original paper
and derive a simple  criterion for the  existence of a single bound 
charged-biexciton state.
\end{abstract}

\bigskip

In his Comment, Varga has used  a AAB-type three-body and 
AAABB-type five-body system to model the charged exciton
and charged biexciton respectively. 
These excitonic systems  have  been modelled as an AB type 
Coulomb problem in our work \cite{thil}. This is done by 
formulating the two-body method (Eqs (15-19) in our original 
paper) so that a radial Schrodinger equation is obtained in which
the potential is not the real two-body potential,
but the average potential of a three- or five-body system.
With the use of an effective mass and dielectric constant,
the resulting radial equation approximates the problem 
of an exciton complex interacting
with holes and electrons in the valence and conducting 
bands respectively. 
K. Varga in his  Comment has  critized
this two-body approach and questioned the
validity of our results obtained for the charged-biexciton
system.

We have modelled the charged-biexciton
as a two-body system (biexciton + e or h)  - justifying 
calculations based on the analytical expressions derived
for the charged exciton (exciton + e or h).
The biexciton is  conceptualized as a 
single hydrogenic composite system;
the advantage being that experimental estimates of 
the biexciton binding energy (ref. 
19,20 of our paper) can be used to estimate energies of 
the charged-biexciton complex. 
Due to the semi-empirical nature of our
charged-biexciton model, we firmly disagree with the 
Comment that an over simplified
model has been used in our work.

Varga's Comment uses a  stochastic variational method
based on correlated Guassian basis sets \cite{v1}. 
Unlike  the semi-empirical charged biexciton model 
used in our work,  this method  treats excitonic systems
in solid materials just like atomic systems (e.g. $H_{2}^+$ and 
$H_{3}^+$) which  casts doubts on the values of binding 
energies obtained for the charged-biexciton. 
Due to the lack of analytical forms \cite{bas} for the Bloch 
functions, it is almost impossible to determine the correct 
dielectric function and hence
the correct Coulomb potential between the electrons and holes in 
a  charged biexciton. The choice of a suitable variational
wavefunction to calculate the 
binding energy of a highly complex system like the 
charged-biexciton is thus a difficult one to make. 
This constrast with our heuristic approach\cite{thil}
in which most of the properties of the solid
are coalesced into the effective masses and dielectric contant 
of the interacting  electron and hole.
Thus the 
sensitivity of assumed forms of variational functions on 
calculated results as well as the role of the geometrical 
structure \cite{chap} and hidden symmetry \cite{jjp} in 
stabilizing the charged exciton complex have to be fully explored
before a definitive statement about the stability of the 
charged-biexciton is made.

In this respect, we point out a misleading statement
in our original paper\cite{thil}. The sentence 
that the "charged-biexciton complex is stable for any value 
of the mass ratio" which  appears under "Results and Discussion" 
is somewhat misleading. It should be noted that eq.25 
which specifies  the {\it upper bound} in the ratio of
binding energy of the charged exciton to charged biexciton, 
(stated in the paragraph preceding Eq.25)  
should correctly be an inequality ($\leq$
instead of $=$ in Eq.25). Hence Eq.25 does not ensure the 
existence of  {\it at least one bound state}. A simple and 
efficient criterion for the existence of a single bound state
can be obtained using eq.24:
\begin{equation}
Eb_{X_3^{-}} = 
({M_{X_3^{-}} \over {M_{R'}}} {\epsilon_{2ex}^{2} 
\over \epsilon_{X_3^{-}}^2} -1) Eb_{2ex},
\label{e1}
\end{equation}

Using relations derived for a
biexciton \cite{jai},
${M_{R'}} \approx {2 \over 3} \mu_{ex}$ and  
$\epsilon_{2ex} \approx {\sqrt{2} \over 4-\sqrt{2}} \epsilon_{ex} $,
and    eq.(\ref{e1}), we get a sufficient
condition for the existence of charged-biexciton states:

\begin{equation}
\gamma_i > 2.23 \; \; \; {\rm where} \; \; 
\gamma_i =  {M_{X_3^{i}} \over \mu_{ex}} 
{{\epsilon_{ex}}^2 \over {\epsilon_{X_3^{i}}}^2} 
\label{e2}
\end{equation}
where i=+ or -,  depending on the charge of the 
charged-biexciton. 
$\gamma_{+} > \gamma_{-}$ due to the larger mass of 
holes and thus the positively charged-biexciton 
is relatively  stable compared to its negatively charged 
counterpart. Unlike in three dimensional
systems, two-body bound states  have been 
shown \cite{p1,p2} to occur even for arbitrarily small 
attractive potentials in two and one dimensional systems.
This fact was well demonstrated by 
recent experiments \cite{mag} which showed that 
a finite exciton binding energy remains present up to 
very high free-carrier concentrations.
Extending the results of these works
to eq.(\ref{e2}), we can 
expect at least a single bound state of the charged-biexciton 
(${X_3^{-}}$ and ${X_3^{+}}$) for small values of $\sigma$. This is 
in contradiction to the results in Varga's Comment.

\end{document}